\documentclass[10pt,conference]{IEEEtran}

\ifx\pdfoutput\undefined
\usepackage{graphicx}
\graphicspath{{eps_figs/}}
\else
\usepackage[pdftex]{graphicx}
\graphicspath{{pdf_figs/}}
\fi

\usepackage{latexsym,amssymb,amsmath,cite,ifthen,afterpage}

\newcommand{\Fig}[1]{Fig.~\ref{#1}}

\newcommand{\eqdef}{\stackrel{\scriptscriptstyle\bigtriangleup}{=} }

\newcommand{\R}{\mathbb{R}}

\newcommand{\calX}{\mathcal{X}}

\newcommand{\calU}{\mathcal{U}}

\newcounter{examplecntr}
{\begin{trivlist}\small\item[]\refstepcounter{examplecntr}%
 {\bfseries Example~\theexamplecntr%
  \ifthenelse{\equal{#1}{}}{}{ (#1)}.
}}%
{\end{trivlist}}

\newcounter{theoremcntr}
{\begin{trivlist}\item[]\refstepcounter{theoremcntr}%
{\bfseries Theorem~\thetheoremcntr%
  \ifthenelse{\equal{#1}{}}{}{ (#1)}.
}}%
{\hfill$\Box$\end{trivlist}}

\newcommand{\pos}[2]{\makebox(0,0)[#1]{#2}}

\begin{document}
\DeclareGraphicsExtensions{.pdf}

\title{Partition Function of the Ising Model\linebreak 
via Factor Graph Duality}


\author{\IEEEauthorblockN{Mehdi Molkaraie}
\IEEEauthorblockA{University of Waterloo \\
Dept.\ of Statistics \& Actuarial Science \\
Waterloo, Canada N2L 3G1 \\
mmolkaraie@uwaterloo.ca}
\and
\IEEEauthorblockN{Hans-Andrea Loeliger}
\IEEEauthorblockA{ETH Zurich\\
Dept.\ of Information Technology \& Electrical Engineering\\ 
8092 Z\"urich, Switzerland\\ 
loeliger@isi.ee.ethz.ch}
}

\maketitle

\begin{abstract}
The partition function of a factor graph and the partition function
of the dual factor graph are related to each other by the normal 
factor graph duality theorem. We apply this result to the classical problem of 
computing the partition function
of the Ising model. In the one-dimensional case, we thus obtain an 
alternative derivation
of the (well-known) analytical solution.  In the two-dimensional case,
we find that Monte Carlo methods are much more efficient on the dual graph 
than on the original graph, especially at low temperature.

\end{abstract}

\section{Introduction}
\label{sec:Introduction}

We consider the problem of computing the partition 
function of one-dimensional (1D) and finite-size 
two-dimensional (2D) Ising models.
In particular, we propose a method
to compute the partition function of finite-size 2D Ising 
models at low temperature 
by performing
Markov chain Monte Carlo methods on the dual factor 
graph.

The problem setup is as follows. 
Let $X_1, X_2, \ldots, X_N$ be random variables, each
taking its values in $\calX = \{0, 1\}$. (In statistical
physics, variables are usually considered as particles which take on
two possible states, e.g., spin up $\uparrow$ and spin 
down $\downarrow$). 
Let $x_i$ represent a
possible realization of $X_i$ and let $x$ stand for 
a configuration $(x_1, x_2, \ldots, x_N)$. 

\begin{figure}
\setlength{\unitlength}{0.93mm}
\centering
\begin{picture}(76,64)(0,0)
\small
\put(0,60){\framebox(4,4){$=$}}
\put(4,62){\line(1,0){8}}
\put(12,60){\framebox(4,4){}}
\put(16,62){\line(1,0){8}}
\put(24,60){\framebox(4,4){$=$}}
\put(28,62){\line(1,0){8}}
\put(36,60){\framebox(4,4){}}
\put(40,62){\line(1,0){8}}
\put(48,60){\framebox(4,4){$=$}}
\put(52,62){\line(1,0){8}}
\put(60,60){\framebox(4,4){}}
\put(64,62){\line(1,0){8}}
\put(72,60){\framebox(4,4){$=$}}
\put(2,54){\line(0,1){6}}
\put(0,50){\framebox(4,4){}}
\put(2,50){\line(0,-1){6}}
\put(26,54){\line(0,1){6}}
\put(24,50){\framebox(4,4){}}
\put(26,50){\line(0,-1){6}}
\put(50,54){\line(0,1){6}}
\put(48,50){\framebox(4,4){}}
\put(50,50){\line(0,-1){6}}
\put(74,54){\line(0,1){6}}
\put(72,50){\framebox(4,4){}}
\put(74,50){\line(0,-1){6}}
\put(0,40){\framebox(4,4){$=$}}
\put(4,42){\line(1,0){8}}
\put(12,40){\framebox(4,4){}}
\put(16,42){\line(1,0){8}}
\put(24,40){\framebox(4,4){$=$}}
\put(28,42){\line(1,0){8}}
\put(36,40){\framebox(4,4){}}
\put(40,42){\line(1,0){8}}
\put(48,40){\framebox(4,4){$=$}}
\put(52,42){\line(1,0){8}}
\put(60,40){\framebox(4,4){}}
\put(64,42){\line(1,0){8}}
\put(72,40){\framebox(4,4){$=$}}
\put(2,34){\line(0,1){6}}
\put(0,30){\framebox(4,4){}}
\put(2,30){\line(0,-1){6}}
\put(26,34){\line(0,1){6}}
\put(24,30){\framebox(4,4){}}
\put(26,30){\line(0,-1){6}}
\put(50,34){\line(0,1){6}}
\put(48,30){\framebox(4,4){}}
\put(50,30){\line(0,-1){6}}
\put(74,34){\line(0,1){6}}
\put(72,30){\framebox(4,4){}}
\put(74,30){\line(0,-1){6}}
\put(0,20){\framebox(4,4){$=$}}
\put(4,22){\line(1,0){8}}
\put(12,20){\framebox(4,4){}}
\put(16,22){\line(1,0){8}}
\put(24,20){\framebox(4,4){$=$}}
\put(28,22){\line(1,0){8}}
\put(36,20){\framebox(4,4){}}
\put(40,22){\line(1,0){8}}
\put(48,20){\framebox(4,4){$=$}}
\put(52,22){\line(1,0){8}}
\put(60,20){\framebox(4,4){}}
\put(64,22){\line(1,0){8}}
\put(72,20){\framebox(4,4){$=$}}
\put(2,14){\line(0,1){6}}
\put(0,10){\framebox(4,4){}}
\put(2,10){\line(0,-1){6}}
\put(26,14){\line(0,1){6}}
\put(24,10){\framebox(4,4){}}
\put(26,10){\line(0,-1){6}}
\put(50,14){\line(0,1){6}}
\put(48,10){\framebox(4,4){}}
\put(50,10){\line(0,-1){6}}
\put(74,14){\line(0,1){6}}
\put(72,10){\framebox(4,4){}}
\put(74,10){\line(0,-1){6}}
\put(0,0){\framebox(4,4){$=$}}
\put(4,2){\line(1,0){8}}
\put(12,0){\framebox(4,4){}}
\put(16,2){\line(1,0){8}}
\put(24,0){\framebox(4,4){$=$}}
\put(28,2){\line(1,0){8}}
\put(36,0){\framebox(4,4){}}
\put(40,2){\line(1,0){8}}
\put(48,0){\framebox(4,4){$=$}}
\put(52,2){\line(1,0){8}}
\put(60,0){\framebox(4,4){}}
\put(64,2){\line(1,0){8}}
\put(72,0){\framebox(4,4){$=$}}
\put(8,63){\pos{bc}{$X_1$}}
\put(32,63){\pos{bc}{$X_2$}}
\put(56,63){\pos{bc}{$X_3$}}
\end{picture}
\caption{\label{fig:2DGrid}
Factor graph of a 2D Ising model. The unlabeled boxes
represent factors as in~(\ref{eqn:Kappa}).}
\end{figure}
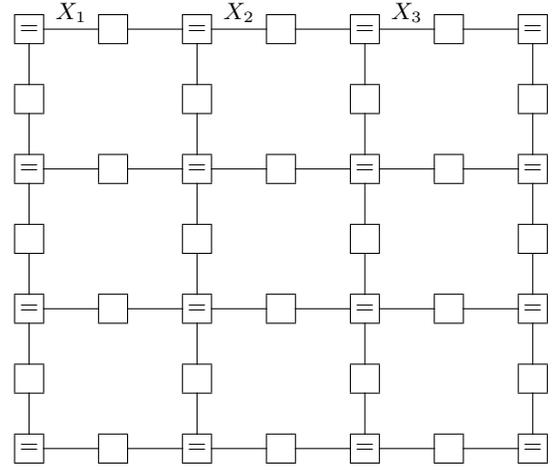

In 1D Ising models, as shown in 
Figs.~\ref{fig:1DIsing} and \ref{fig:1DIsingFree}, 
variables (particles) are considered as a chain of sites on a line. In 2D
Ising models, variables are arranged on the sites of a 2D lattice, 
as depicted in Fig.~\ref{fig:2DGrid}.

We assume that only \emph{adjacent (nearest neighbor)} variables
interact with each other. 
In the absence of an external field, we define 
the energy of a configuration $x$ as~\cite{Huang:87}

\begin{equation}
\label{eqn:Hamiltonian}
E(x) \eqdef -\!\!\!\!\sum_{\text{$k,\ell$ adjacent}}\!\!\!\!J_{k, \ell}
\big([x_k = x_{\ell}] - [x_k \ne x_{\ell}]\big) 
\end{equation}
where the sum runs over all the (unordered) adjacent pairs $(k,\ell)$
and $[\cdot]$ denotes the Iverson 
bracket~\cite[p.\ 24]{GKP:89}, which evaluates to one if the condition in 
the bracket is satisfied and to zero otherwise.

The real coupling parameter $J_{k, \ell}$, controls the strength of
the interaction between $(x_k, x_{\ell})$.
If $J_{k,\ell} > 0$, the model is known 
as a ferromagnetic Ising model. The model is called antiferromagnetic
if $J_{k,\ell} < 0$, 
see~\cite{Cipra:87,Huang:87}. If the couplings can be both positive
or negative (e.g., chosen at random 
according to some
distribution) the model is usually known as an Ising spin 
glass, see~\cite[Chapter 12]{MezMon:92}. 

In thermal equilibrium, the probability of a configuration $x$, is
given by the Boltzmann distribution, defined as~\cite{Huang:87}
\begin{equation} 
\label{eqn:Prob}
p_{\text{B}}(x) \eqdef \frac{e^{-\beta E(x)}}{Z} 
\end{equation}
where $Z$ is the \emph{partition function} (normalization constant) and
$\beta \eqdef \frac{1}{k_{\text{B}}T}$, where
temperature is denoted by $T$, and $k_{\text{B}}$ is the Boltzmann 
constant.

In the rest of this paper, we will assume $\beta = 1$. With this 
assumption,
small values of $|J|$ correspond to models at high temperature, 
and large values of $|J|$ to models at low temperature.

The Helmholtz free energy is defined 
as
\begin{equation} 
\label{eqn:FreeEnergy}
F_{\text{H}} \eqdef -\ln Z,
\end{equation}
see~\cite{Huang:87}.

For each adjacent pair $(k,\ell)$, let
\begin{equation}
\label{eqn:Kappa}
\kappa_{k, \ell}(x_k, x_{\ell}) = e^{J_{k, \ell}([x_k = x_{\ell}] - [x_k \ne x_{\ell}])}
\end{equation}
and let $f: \calX^N \rightarrow \R$ be 
\begin{equation} 
\label{eqn:factorF}
f(x) =  \prod_{\text{$k,\ell$ adjacent}} \kappa_{k, \ell}(x_k, x_{\ell})
\end{equation}
where the product runs over all the (unordered) adjacent pairs $(k,\ell)$.

We are interested in computing the partition function
\begin{IEEEeqnarray}{rCl}
Z & = &  \sum_{x\in \calX^N} e^{-E(x)} \label{eqn:ZSum1}\\
 & = & \sum_{x\in \calX^N}\prod_{\text{$k,\ell$ adjacent}} e^{J_{k, \ell}
\big([x_k = x_{\ell}] - [x_k \ne x_{\ell}]\big)} \label{eqn:ZSum2} \\
 & = & \sum_{x \in \calX^N}\prod_{\text{$k,\ell$ adjacent}} \kappa_{k, \ell}(x_k,x_\ell) \label{eqn:ZKappa}\\
   & = & \sum_{x \in \calX^N} f(x). \label{eqn:ZKappa2}
\end{IEEEeqnarray}


\begin{figure}
\setlength{\unitlength}{1.15mm}
\centering
\begin{picture}(52,8)(0,0)
\small
\put(-8,2){\line(1,0){8}}
\put(-4,3){\pos{bc}{$X_1$}}
\put(0,0){\framebox(4,4){}}
\put(8,3){\pos{bc}{$X_2$}}
\put(4,2){\line(1,0){8}}
\put(12,0){\framebox(4,4){}}
\put(20,3){\pos{bc}{$X_3$}}
\put(16,2){\line(1,0){8}}
\put(24,0){\framebox(4,4){}}
\put(32,3){\pos{bc}{$X_4$}}
\put(28,2){\line(1,0){8}}
\put(36,0){\framebox(4,4){}}
\put(44,3){\pos{bc}{$X_5$}}
\put(40,2){\line(1,0){8}}
\put(48,0){\framebox(4,4){}}
\put(52,2){\line(1,0){8}}
\put(-8,2){\line(0,1){4.32}}
\put(60,2){\line(0,1){4.32}}
\put(-8,6.32){\line(1,0){68}}
\end{picture}
\caption{\label{fig:1DIsing}
Factor graph of a 1D Ising model with $N = 5$ and 
with periodic boundary conditions. The
solid boxes represent factors as in~(\ref{eqn:1DKernel}), and 
edges represent the variables.}
\centering
\begin{picture}(40,8)(0,0)
\small
\put(-11,1){\framebox(2,2){}}
\put(-8,2){\line(1,0){8}}
\put(-4,3){\pos{bc}{$X_1$}}
\put(0,0){\framebox(4,4){}}
\put(8,3){\pos{bc}{$X_2$}}
\put(4,2){\line(1,0){8}}
\put(12,0){\framebox(4,4){}}
\put(20,3){\pos{bc}{$X_3$}}
\put(16,2){\line(1,0){8}}
\put(24,0){\framebox(4,4){}}
\put(32,3){\pos{bc}{$X_4$}}
\put(28,2){\line(1,0){8}}
\put(36,0){\framebox(4,4){}}
\put(44,3){\pos{bc}{$X_5$}}
\put(40,2){\line(1,0){8}}
\put(49,1){\framebox(2,2){}}
\end{picture}
\caption{\label{fig:1DIsingFree}
Factor graph of a 1D Ising model with $N = 5$ and 
with free boundary conditions.
The normal-size
boxes represent factors as in~(\ref{eqn:1DKernel}),
and the two small boxes represent constant factors.}
\end{figure}

In 1D Ising models, $f$ has a cycle-free factor graph 
representation and
$Z$, as in (\ref{eqn:ZKappa2}), can be computed 
directly by sum-product message passing~\cite{KFL:01, Lg:ifg2004},
which (in this context) coincides with the transfer matrix method in 
statistical physics~\cite{Baxter:07},~\cite[Chapter 5]{Yeo:92}.

In 2D Ising models with constant coupling and in the absence of an 
external field, the exact value of $Z$ in thermodynamic limits 
(for $N \to \infty$) was found by 
Onsager~\cite{Onsager:44}.

For finite-size 2D Ising models with arbitrary coupling, 
estimates of the partition function can be computed by Markov chain Monte 
Carlo methods~\cite{Neal:proinf1993r, MK:mct1998, LoMo:IT2013}.
At high temperatures, the Boltzmann 
distribution~(\ref{eqn:Prob})
tends to a uniform distribution and Monte Carlo methods 
generally work very well.
At low temperatures, however, variables have long-range interactions; 
Monte Carlo methods are plagued by slow and erratic convergence, 
and may break down completely.

In this paper, we consider using the dual factor graph~\cite{Forney:01, AY:2011, FV:2011}
to compute 
(or to estimate) the partition function of the Ising model.
In the one-dimensional case, we will thus obtain an alternative derivation
of the (well-known) analytical solution. In the two-dimensional case,
we find that Monte Carlo methods are much more efficient (due to much 
faster mixing)
on the dual graph than on the original graph, especially at low temperature.

The paper is structured as follows.
In Section~\ref{sec:NFGD}, we recall the
construction of the dual factor graph and the factor graph duality theorem. 
In Section~\ref{sec:1DIsing}, we use this theorem to obtain the partition 
function of 1D Ising models.
In Section~\ref{sec:2DIsing}, we discuss the dual factor graph of 2D Ising 
models, which we then use for the numerical simulations in 
Section~\ref{sec:NumExpr}.

\section{Partition Function via Factor Graph Duality}
\label{sec:NFGD}

The factorization of a function, as in~(\ref{eqn:factorF}), can be 
represented
by a Forney factor graph. 
The nodes in a Forney factor graph represent the factors 
and the edges (or half-edges, which are connected to only one node) 
represent the variables.
The edge (or half-edge) that represents some variable $x$, is
connected to the node representing the factor $\kappa$,
if and only if $x$ is an argument of $\kappa$, see~\cite{Forney:01, Lg:ifg2004}.

In the factor graphs that we study in this paper, all variables are
binary and there are no half edges.
Starting from such a factor graph, we can obtain its dual by 
replacing each variable $x$ with its dual (frequency) 
variable $\tilde x$, each 
factor $\kappa$ with its Fourier transform $\nu$, and each equality 
constraint with an XOR factor~\cite{Forney:01, AY:2011}.

For binary variables $\tilde x_1, \tilde x_2, \ldots, x_k$, the 
XOR factor is defined as
\begin{equation} 
\label{eqn:XOR}
g(\tilde x_1, \tilde x_2, \ldots, \tilde x_k) \eqdef 
[\tilde x_1 \oplus \tilde x_2 \oplus \ldots \oplus \tilde x_k=0]
\end{equation}
where $\oplus$ denotes addition modulo 2.

Note that, in general, factors in the dual Forney graph can be
negative or even complex-valued~\cite{Brace:1999,MoLo:ITW2012}.

Random variables in the dual domain are denoted by $\tilde X$,
which also take their values in $\calX$.
In the dual Forney factor graph, we denote the partition function
by $Z_d$, and the number of edges by $E$. 
In such a set-up, according to the factor graph duality theorem
\cite[Theorem 2]{AY:2011}, 
\begin{equation}
\label{eqn:NDual}
Z_d = |\calX^E|Z
\end{equation}

Therefore, one alternative method to compute $Z$, is to first
compute the partition function of the dual factor graph $Z_d$, and then
apply~(\ref{eqn:NDual}), see~\cite{AY:2011, FV:2011}. This procedure is 
particularly 
useful, in the cases that computing the partition function can 
be performed more efficiently in the dual domain.
In this paper, we show that this is indeed the case for the Ising 
model, especially at low temperature.

\section{1D Ising Models}
\label{sec:1DIsing}

We consider a 1D Ising model with size $N$, with $N$ binary (i.e.,~$\{0, 1\}$-valued)
variables, and with \emph{periodic} boundary conditions
(i.e., $X_{N+1} = X_1$).

For $1 \le \ell \le N$, we have
\begin{equation} 
\label{eqn:1DKernel}
\kappa_\ell(x_\ell, x_{\ell+1}) = \left\{ \begin{array}{ll}
     e^{J_{\ell}}, & \text{if $x_\ell = x_{\ell+1}$} \\
     e^{-J_{\ell}}, & \text{if $x_\ell \neq x_{\ell+1}$}
  \end{array} \right.
\end{equation}

Therefore,
\begin{equation}
Z = \sum_{x \in \calX^N}\prod_{\ell = 1}^{N}\kappa_\ell(x_\ell, x_{\ell+1})
\end{equation}

The corresponding factor graph of $f$ is shown in Fig.~\ref{fig:1DIsing}.
Note that, in order to create periodic boundary conditions, we have 
simply joined the two ends of the factor graph.

To construct the dual factor graph, each 
factor~(\ref{eqn:1DKernel}) is replaced by 
its 2D discrete Fourier transform (DFT),
where
the 2D DFT  $\nu(\tilde x_1, \tilde x_2)$ of $\kappa(x_1, x_2)$, is defined as

\begin{equation} 
\nu(\tilde x_1, \tilde x_2) \eqdef 
\sum_{x_1\in \calX}\sum_{x_2 \in \calX} \kappa(x_1,x_2)
e^{-i2\pi(x_1\tilde x_1 + x_2\tilde x_2)/|\calX|}
\end{equation}
where $i = \sqrt{-1}$. 

Therefore in the dual factor graph, each factor
$\nu_\ell(\tilde x_\ell, \tilde x_{\ell+1})$ has the following form
\begin{equation} 
\label{eqn:1DKernelDual}
\nu_\ell(\tilde x_\ell, \tilde x_{\ell+1}) = \left\{ \begin{array}{ll}
     4\cosh J_\ell, & \text{if $\tilde x_\ell = \tilde x_{\ell+1} = 0$} \\
     4\sinh J_\ell, & \text{if $\tilde x_\ell = \tilde x_{\ell+1} = 1$} \\
     0, & \text{otherwise.}
  \end{array} \right.
\end{equation}

Computing $Z_d$ is now straightforward since all the factors 
in~(\ref{eqn:1DKernelDual}) are diagonal.
We conclude that
\begin{IEEEeqnarray}{rCl}
Z_d & = & \sum_{\tilde x \in \calX^N}\prod_{\ell = 1}^{N}\nu_\ell(\tilde x_\ell, \tilde x_{\ell+1}) \\
       & = & 4^{N} \big(\prod_{\ell = 1}^N\cosh J_\ell + \prod_{\ell=1}^N\sinh J_\ell\big).
\end{IEEEeqnarray}

The number of edges in the dual factor graph is $N$.
Using the factor graph duality theorem~(\ref{eqn:NDual}), we 
obtain
\begin{equation} 
Z = 2^{N} \big(\prod_{\ell=1}^N\cosh J_\ell + \prod_{\ell=1}^N\sinh J_\ell\big).
\end{equation}

Finally, we state (without proof) that computing
$Z$ of a 1D Ising model with size $N$ and with \emph{free} 
boundary conditions is also straightforward.
After introducing two constant
``dummy" factors at the two ends of the factor graph,
as illustrated in Fig.~\ref{fig:1DIsingFree}, we can
directly apply the factor graph duality theorem to compute the
partition function as
\begin{equation}
Z = 2\prod_{\ell = 1}^{N-1}(2\cosh J_\ell)
\end{equation}
see~\cite[Chapter 2]{Baxter:07},\cite[Chapter 5]{Yeo:92}.


\newcommand{\drawgrid}{%
\begin{picture}(76,64)(0,0)
\small
\put(0,60){\framebox(4,4){+}}
\put(4,62){\line(1,0){8}}
\put(12,60){\framebox(4,4){=}}
\put(16,62){\line(1,0){8}}
\put(24,60){\framebox(4,4){$+$}}
\put(28,62){\line(1,0){8}}
\put(36,60){\framebox(4,4){=}}
\put(40,62){\line(1,0){8}}
\put(48,60){\framebox(4,4){$+$}}
\put(52,62){\line(1,0){8}}
\put(60,60){\framebox(4,4){=}}
\put(64,62){\line(1,0){8}}
\put(72,60){\framebox(4,4){$+$}}
\put(2,54){\line(0,1){6}}
\put(0,50){\framebox(4,4){=}}
\put(2,50){\line(0,-1){6}}
\put(26,54){\line(0,1){6}}
\put(24,50){\framebox(4,4){=}}
\put(26,50){\line(0,-1){6}}
\put(50,54){\line(0,1){6}}
\put(48,50){\framebox(4,4){=}}
\put(50,50){\line(0,-1){6}}
\put(74,54){\line(0,1){6}}
\put(72,50){\framebox(4,4){=}}
\put(74,50){\line(0,-1){6}}
\put(0,40){\framebox(4,4){$+$}}
\put(4,42){\line(1,0){8}}
\put(12,40){\framebox(4,4){=}}
\put(16,42){\line(1,0){8}}
\put(24,40){\framebox(4,4){$+$}}
\put(28,42){\line(1,0){8}}
\put(36,40){\framebox(4,4){=}}
\put(40,42){\line(1,0){8}}
\put(48,40){\framebox(4,4){$+$}}
\put(52,42){\line(1,0){8}}
\put(60,40){\framebox(4,4){=}}
\put(64,42){\line(1,0){8}}
\put(72,40){\framebox(4,4){$+$}}
\put(2,34){\line(0,1){6}}
\put(0,30){\framebox(4,4){=}}
\put(2,30){\line(0,-1){6}}
\put(26,34){\line(0,1){6}}
\put(24,30){\framebox(4,4){=}}
\put(26,30){\line(0,-1){6}}
\put(50,34){\line(0,1){6}}
\put(48,30){\framebox(4,4){}}
\put(50,30){\line(0,-1){6}}
\put(74,34){\line(0,1){6}}
\put(72,30){\framebox(4,4){=}}
\put(74,30){\line(0,-1){6}}
\put(0,20){\framebox(4,4){$+$}}
\put(4,22){\line(1,0){8}}
\put(12,20){\framebox(4,4){=}}
\put(16,22){\line(1,0){8}}
\put(24,20){\framebox(4,4){$+$}}
\put(28,22){\line(1,0){8}}
\put(36,20){\framebox(4,4){=}}
\put(40,22){\line(1,0){8}}
\put(48,20){\framebox(4,4){$+$}}
\put(52,22){\line(1,0){8}}
\put(60,20){\framebox(4,4){=}}
\put(64,22){\line(1,0){8}}
\put(72,20){\framebox(4,4){$+$}}
\put(2,14){\line(0,1){6}}
\put(0,10){\framebox(4,4){=}}
\put(2,10){\line(0,-1){6}}
\put(26,14){\line(0,1){6}}
\put(24,10){\framebox(4,4){=}}
\put(26,10){\line(0,-1){6}}
\put(50,14){\line(0,1){6}}
\put(48,10){\framebox(4,4){=}}
\put(50,10){\line(0,-1){6}}
\put(74,14){\line(0,1){6}}
\put(72,10){\framebox(4,4){=}}
\put(74,10){\line(0,-1){6}}
\put(0,0){\framebox(4,4){$+$}}
\put(4,2){\line(1,0){8}}
\put(12,0){\framebox(4,4){=}}
\put(16,2){\line(1,0){8}}
\put(24,0){\framebox(4,4){$+$}}
\put(28,2){\line(1,0){8}}
\put(36,0){\framebox(4,4){=}}
\put(40,2){\line(1,0){8}}
\put(48,0){\framebox(4,4){$+$}}
\put(52,2){\line(1,0){8}}
\put(60,0){\framebox(4,4){=}}
\put(64,2){\line(1,0){8}}
\put(72,0){\framebox(4,4){$+$}}
\end{picture}
}

\begin{figure}
\setlength{\unitlength}{0.93mm}
\centering
\begin{picture}(76,64)(0,0)
\small
\put(0,60){\framebox(4,4){+}}
\put(4,62){\line(1,0){8}}
\put(12,60){\framebox(4,4){}}
\put(16,62){\line(1,0){8}}
\put(24,60){\framebox(4,4){$+$}}
\put(28,62){\line(1,0){8}}
\put(36,60){\framebox(4,4){}}
\put(40,62){\line(1,0){8}}
\put(48,60){\framebox(4,4){$+$}}
\put(52,62){\line(1,0){8}}
\put(60,60){\framebox(4,4){}}
\put(64,62){\line(1,0){8}}
\put(72,60){\framebox(4,4){$+$}}
%
\put(2,54){\line(0,1){6}}
\put(0,50){\framebox(4,4){}}
\put(2,50){\line(0,-1){6}}
\put(26,54){\line(0,1){6}}
\put(24,50){\framebox(4,4){}}
\put(26,50){\line(0,-1){6}}
\put(50,54){\line(0,1){6}}
\put(48,50){\framebox(4,4){}}
\put(50,50){\line(0,-1){6}}
\put(74,54){\line(0,1){6}}
\put(72,50){\framebox(4,4){}}
\put(74,50){\line(0,-1){6}}
\put(0,40){\framebox(4,4){$+$}}
\put(4,42){\line(1,0){8}}
\put(12,40){\framebox(4,4){}}
\put(16,42){\line(1,0){8}}
\put(24,40){\framebox(4,4){$+$}}
\put(28,42){\line(1,0){8}}
\put(36,40){\framebox(4,4){}}
\put(40,42){\line(1,0){8}}
\put(48,40){\framebox(4,4){$+$}}
\put(52,42){\line(1,0){8}}
\put(60,40){\framebox(4,4){}}
\put(64,42){\line(1,0){8}}
\put(72,40){\framebox(4,4){$+$}}
\put(2,34){\line(0,1){6}}
\put(0,30){\framebox(4,4){}}
\put(2,30){\line(0,-1){6}}
\put(26,34){\line(0,1){6}}
\put(24,30){\framebox(4,4){}}
\put(26,30){\line(0,-1){6}}
\put(50,34){\line(0,1){6}}
\put(48,30){\framebox(4,4){}}
\put(50,30){\line(0,-1){6}}
\put(74,34){\line(0,1){6}}
\put(72,30){\framebox(4,4){}}
\put(74,30){\line(0,-1){6}}
\put(0,20){\framebox(4,4){$+$}}
\put(4,22){\line(1,0){8}}
\put(12,20){\framebox(4,4){}}
\put(16,22){\line(1,0){8}}
\put(24,20){\framebox(4,4){$+$}}
\put(28,22){\line(1,0){8}}
\put(36,20){\framebox(4,4){}}
\put(40,22){\line(1,0){8}}
\put(48,20){\framebox(4,4){$+$}}
\put(52,22){\line(1,0){8}}
\put(60,20){\framebox(4,4){}}
\put(64,22){\line(1,0){8}}
\put(72,20){\framebox(4,4){$+$}}
\put(2,14){\line(0,1){6}}
\put(0,10){\framebox(4,4){}}
\put(2,10){\line(0,-1){6}}
\put(26,14){\line(0,1){6}}
\put(24,10){\framebox(4,4){}}
\put(26,10){\line(0,-1){6}}
\put(50,14){\line(0,1){6}}
\put(48,10){\framebox(4,4){}}
\put(50,10){\line(0,-1){6}}
\put(74,14){\line(0,1){6}}
\put(72,10){\framebox(4,4){}}
\put(74,10){\line(0,-1){6}}
\put(0,0){\framebox(4,4){$+$}}
\put(4,2){\line(1,0){8}}
\put(12,0){\framebox(4,4){}}
\put(16,2){\line(1,0){8}}
\put(24,0){\framebox(4,4){$+$}}
\put(28,2){\line(1,0){8}}
\put(36,0){\framebox(4,4){}}
\put(40,2){\line(1,0){8}}
\put(48,0){\framebox(4,4){$+$}}
\put(52,2){\line(1,0){8}}
\put(60,0){\framebox(4,4){}}
\put(64,2){\line(1,0){8}}
\put(72,0){\framebox(4,4){$+$}}
\put(8,63){\pos{bc}{$\tilde X_1$}}
\put(20,63){\pos{bc}{$\tilde X_2$}}
\put(32,63){\pos{bc}{$\tilde X_3$}}

\end{picture}
\caption{\label{fig:2DGridD}%
The dual Forney factor graph where unlabeled boxes represent
factors as in~(\ref{eqn:2DKernelDual}) and boxes containing 
$+$ symbols represent XOR factors as in~(\ref{eqn:XOR}).}

\setlength{\unitlength}{0.93mm}
\centering
\begin{picture}(76,74)(0,0)
\put(0,0){\drawgrid}
\put(14,64){\line(0,1){2}}
\put(38,64){\line(0,1){2}}
\put(62,64){\line(0,1){2}}
\put(12,66){\framebox(4,4){$$}}
\put(36,66){\framebox(4,4){$$}}
\put(60,66){\framebox(4,4){$$}}
\put(14,44){\line(0,1){2}}
\put(38,44){\line(0,1){2}}
\put(62,44){\line(0,1){2}}
\put(12,46){\framebox(4,4){$$}}
\put(36,46){\framebox(4,4){$$}}
\put(60,46){\framebox(4,4){$$}}
\put(14,24){\line(0,1){2}}
\put(38,24){\line(0,1){2}}
\put(62,24){\line(0,1){2}}
\put(12,26){\framebox(4,4){$$}}
\put(36,26){\framebox(4,4){$$}}
\put(60,26){\framebox(4,4){$$}}
\put(14,4){\line(0,1){2}}
\put(38,4){\line(0,1){2}}
\put(62,4){\line(0,1){2}}
\put(12,6){\framebox(4,4){$$}}
\put(36,6){\framebox(4,4){$$}}
\put(60,6){\framebox(4,4){$$}}
\put(4,52){\line(1,0){2}}
\put(28,52){\line(1,0){2}}
\put(52,52){\line(1,0){2}}
\put(76,52){\line(1,0){2}}
\put(6,50){\framebox(4,4){$$}}
\put(30,50){\framebox(4,4){$$}}
\put(54,50){\framebox(4,4){$$}}
\put(78,50){\framebox(4,4){$$}}
\put(4,32){\line(1,0){2}}
\put(28,32){\line(1,0){2}}
\put(52,32){\line(1,0){2}}
\put(76,32){\line(1,0){2}}
\put(6,30){\framebox(4,4){$$}}
\put(30,30){\framebox(4,4){$$}}
\put(54,30){\framebox(4,4){$$}}
\put(78,30){\framebox(4,4){$$}}
\put(4,12){\line(1,0){2}}
\put(28,12){\line(1,0){2}}
\put(52,12){\line(1,0){2}}
\put(76,12){\line(1,0){2}}
\put(6,10){\framebox(4,4){$$}}
\put(30,10){\framebox(4,4){$$}}
\put(54,10){\framebox(4,4){$$}}
\put(78,10){\framebox(4,4){$$}}
\put(8,63){\pos{bc}{$\tilde X_1$}}
\put(32,63){\pos{bc}{$\tilde X_2$}}
\put(56,63){\pos{bc}{$\tilde X_3$}}
\end{picture}
\caption{\label{fig:2DGridDMod}
The modified dual Forney factor graph where unlabeled boxes represent
factors as in~(\ref{eqn:2DKernelDualM}) and boxes containing 
$+$ symbols represent XOR factors as in~(\ref{eqn:XOR}).}
\end{figure}

\section{Finite-Size 2D Ising Models}
\label{sec:2DIsing}

We consider a 2D Ising model with size $N = m\times m$,
with binary variables (i.e., $\calX = \{0, 1\}$), 
and with factors as in~(\ref{eqn:Kappa}).
The corresponding Forney factor graph with factors as 
in~(\ref{eqn:Kappa}) is shown in~\Fig{fig:2DGrid},
where the boxes labeled ``$=$'' are equality constraints~\cite{Lg:ifg2004}. 

In the dual Forney factor graph, the equality constraints are replaced by 
XOR factors (\ref{eqn:XOR}), and each factor~(\ref{eqn:Kappa}) by its 
2D DFT, which has the following form
\begin{equation} 
\label{eqn:2DKernelDual}
\nu_{k, \ell}(\tilde x_k, \tilde x_\ell) = \left\{ \begin{array}{ll}
     4\cosh J_{k, \ell}, & \text{if $\tilde x_k = \tilde x_\ell = 0$} \\
     4\sinh J_{k, \ell}, & \text{if $\tilde x_k = \tilde x_\ell = 1$} \\
     0, & \text{otherwise.}
  \end{array} \right.
\end{equation}

The corresponding dual Forney factor graph with factors as 
in~(\ref{eqn:2DKernelDual}) is shown in~\Fig{fig:2DGridD}.

\begin{figure}[t!]
\centering
\includegraphics[width=\linewidth]{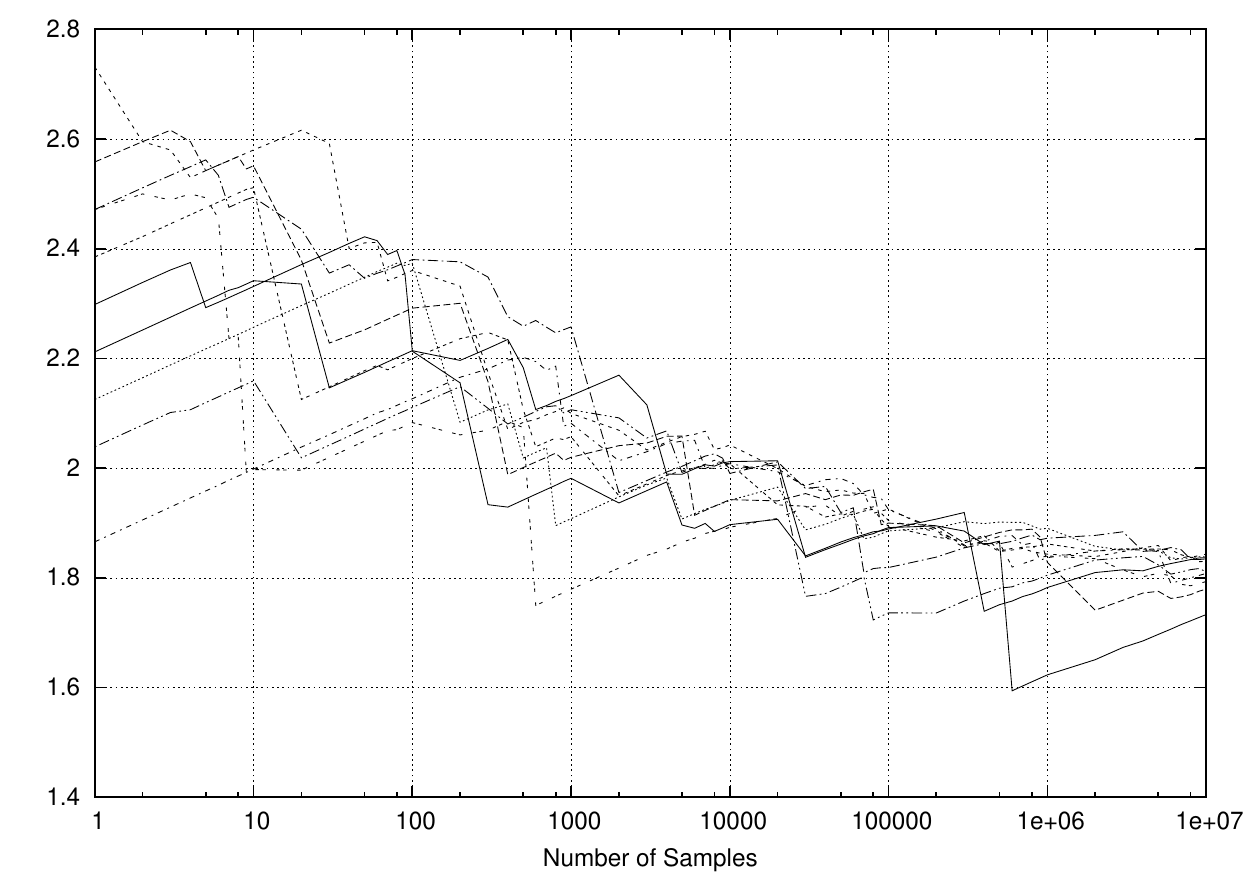}
\caption{\label{fig:FerU}%
Estimated free energy per site vs.\ the number of samples
for a $5\times 5$ ferromagnetic Ising model with 
$J = 0.75$ (low temperature) using Gibbs sampling on the original factor graph.
The plot shows 10 different sample paths.}
\vspace{2mm}
\includegraphics[width=\linewidth]{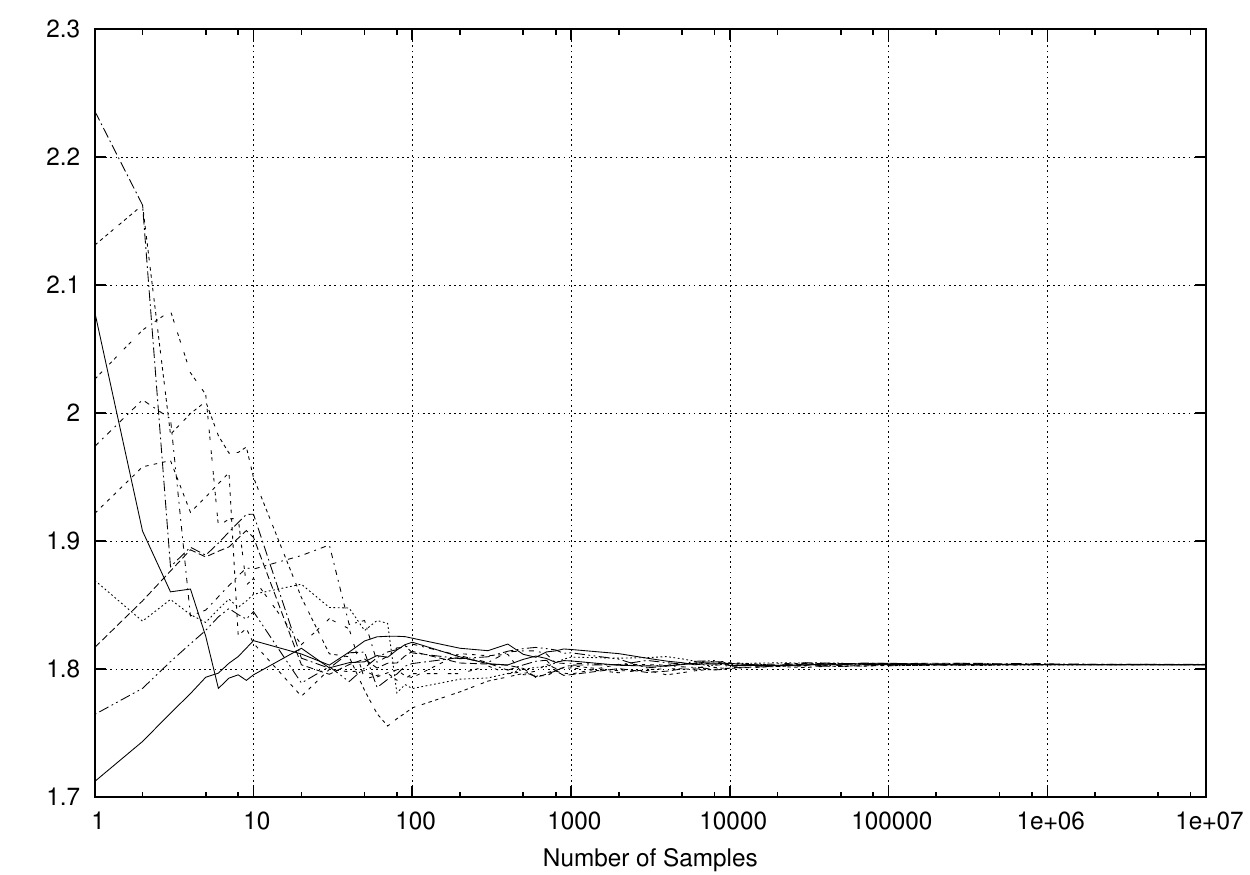}
\caption{\label{fig:FerUD}%
Everything as in Fig.~\ref{fig:FerU}, but on the (modified) dual factor graph.}
\end{figure}

Since all the factors in~(\ref{eqn:2DKernelDual}) are diagonal, 
it is possible to simplify the dual factor 
graph in~\Fig{fig:2DGridD},
to construct the modified dual factor graph depicted in~\Fig{fig:2DGridDMod},
with factors as

\begin{equation} 
\label{eqn:2DKernelDualM}
\nu_k(\tilde x_k) = \left\{ \begin{array}{ll}
     4\cosh J_k, & \text{if $\tilde x_k = 0$} \\
     4\sinh J_k, & \text{if $\tilde x_k = 1$}
  \end{array} \right.
\end{equation}

The corresponding modified dual Forney factor graph with factors as 
in~(\ref{eqn:2DKernelDualM}) is shown in~\Fig{fig:2DGridDMod}.

We are interested in computing the partition 
function, as in~(\ref{eqn:ZKappa}).
In our numerical experiments in
Section~\ref{sec:NumExpr}, we will consider the problem of computing
an estimate of 
the partition function (or equivalently the free energy~(\ref{eqn:FreeEnergy})) 
of 2D Ising 
models (with constant or with spatially varying couplings) 
by Monte Carlo methods as in~\cite{LoMo:IT2013}, on the 
original factor graph with factors as in~(\ref{eqn:Kappa}), and 
on the modified 
dual factor graph with factors as in~(\ref{eqn:2DKernelDualM}).
Monte Carlo methods on the dual graph may be viewed 
as simulating the cycles rather than individual variables. 

\section{Numerical Experiments}
\label{sec:NumExpr}

We apply Monte Carlo methods
to compute the free energy~(\ref{eqn:FreeEnergy}) per 
site, i.e.,  $\frac{1}{N}\log_2 Z$, of 
2D Ising models with size
$N = m\times m$.
Since the value of Z is invariant under the change of
sign of $J$, we will only consider ferromagnetic Ising models.

In Section~\ref{sec:2DConstant}, we consider 2D 
ferromagnetic Ising models, in which the coupling parameter 
$J$ is a positive constant (cf.~Section~\ref{sec:Introduction}). 
For different values of $J$, we compare the
convergence of Gibbs sampling using the 
Ogata-Tanemura method~\cite{LoMo:IT2013, OgTa:eip1981}
and uniform sampling~\cite{MK:mct1998, MoLo:ITW2012} in the original 
factor graph, as in Fig.~\ref{fig:2DGrid}, and in the modified dual factor 
graph, as in Fig.~\ref{fig:2DGridDMod}.
In Section~\ref{sec:2DSpin}, we apply uniform sampling 
on the modified dual factor 
graph to compute the free energy per site of 2D
ferromagnetic Ising models with spatially varying couplings.

\begin{figure}[t!]
\centering
\includegraphics[width=\linewidth]{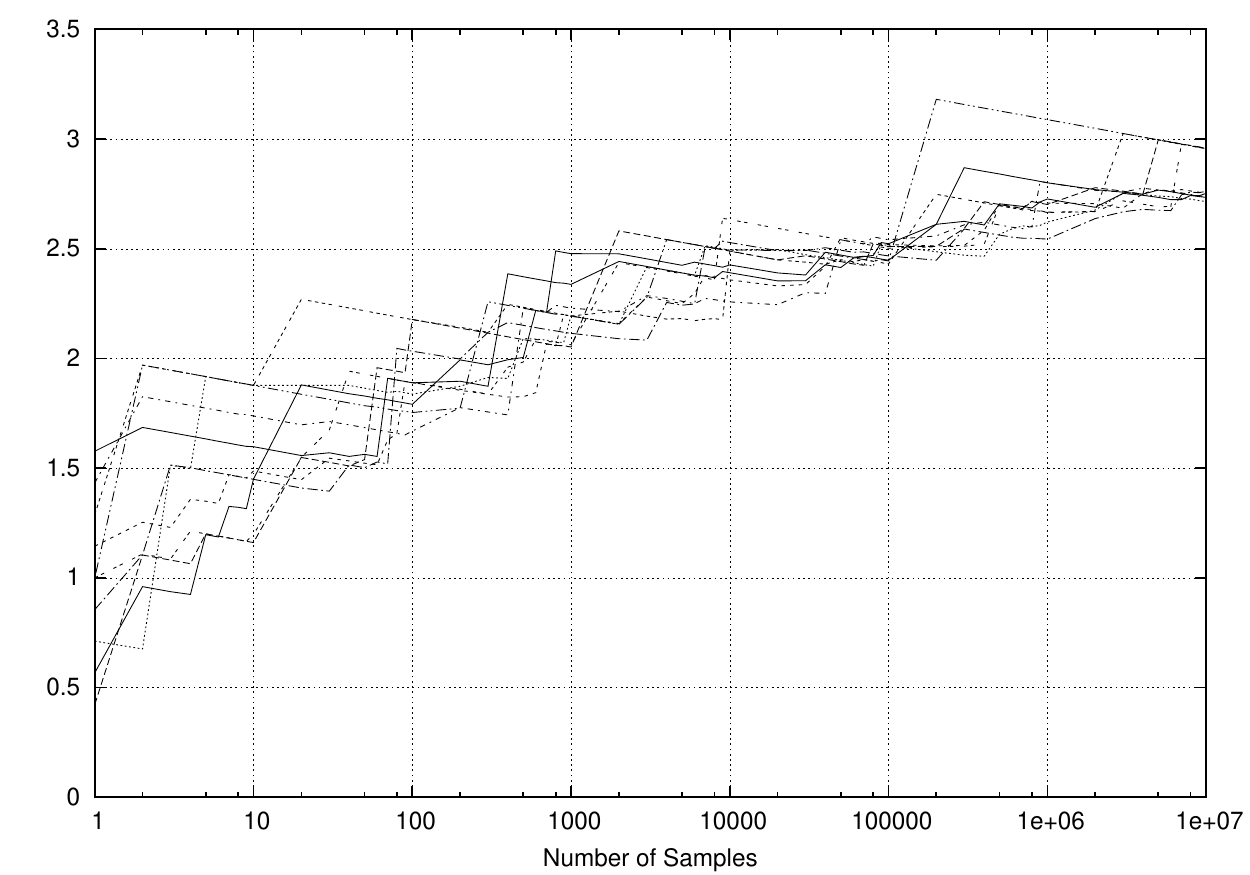}
\caption{\label{fig:FerG}%
Estimated free energy per site vs.\ the number of samples
for a $5\times 5$ ferromagnetic Ising model with $J = 1.25$
(low temperature) using uniform sampling on the original factor graph.
The plot shows 10 different sample paths.}
\vspace{2mm}
\includegraphics[width=\linewidth]{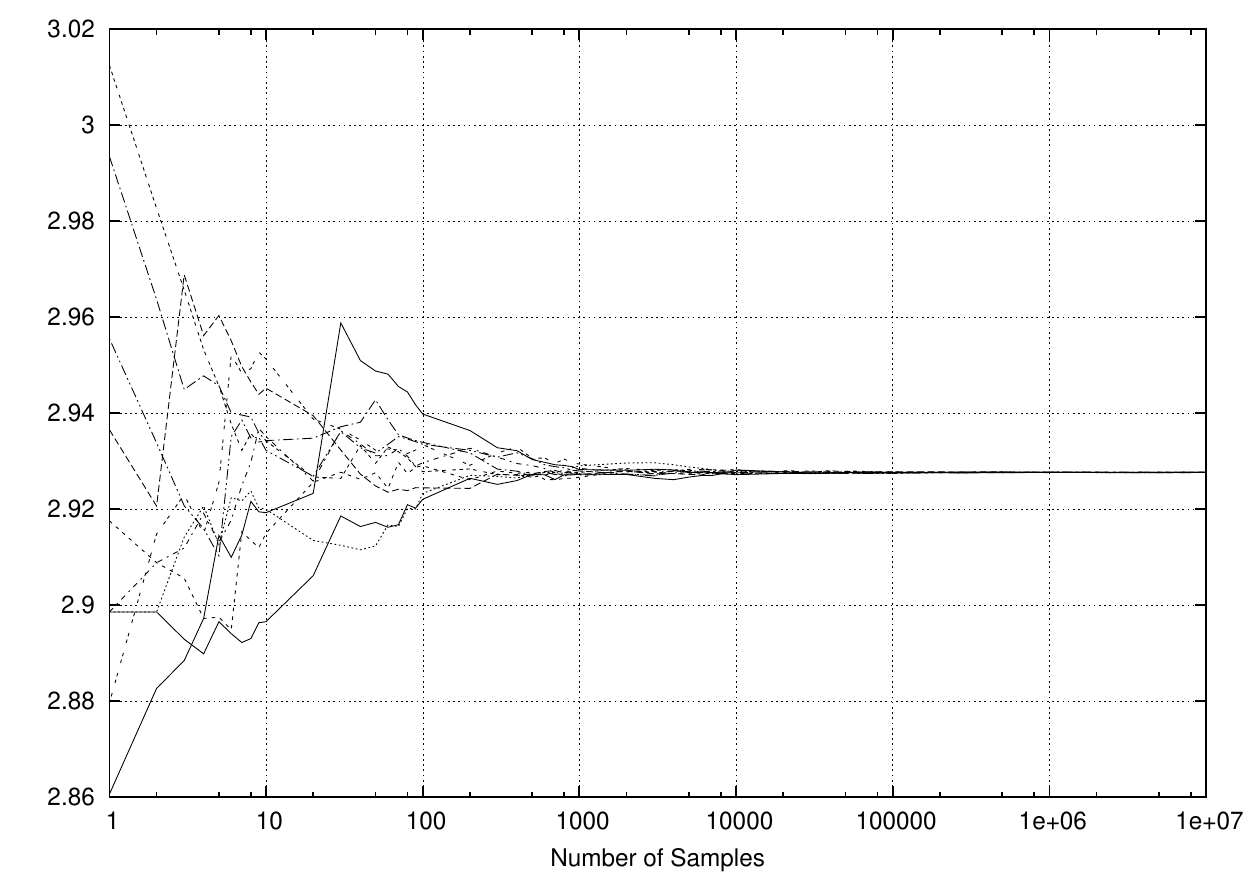}
\caption{\label{fig:FerGD}%
Everything as in Fig.~\ref{fig:FerG}, but on the (modified) dual factor graph.}
\end{figure}

Note that in the modified 
dual factor graph, we can partition the set of random variables $\tilde X$, 
into $\tilde X_A$ and $\tilde X_B$, with the property that the random 
variables in $\tilde X_B$ are 
linear combinations (involving the XOR factors) of the random variables in
$\tilde X_A$. 
Therefore, Monte Carlo methods can be applied directly 
on $\tilde X_A$, then $\tilde X_B$ can be updated
at each iteration according to the new values of $\tilde X_A$. 
In this case, the size of the state space for Monte Carlo
methods only depends on the size of $\tilde X_A$. 

\subsection{2D Ising models with constant coupling}
\label{sec:2DConstant}

We estimate the free energy per 
site, i.e.,  $\frac{1}{N}\log_2 Z$, of 2D ferromagnetic Ising models
with size $N = 5\times 5$ at relatively low and at very low temperatures.
For $J = 0.75$, Figs.~\ref{fig:FerU} and \ref{fig:FerUD} show simulation
results obtained from Gibbs sampling on the original factor graph and on
the modified dual factor graph, respectively. From Fig.~\ref{fig:FerUD}, the 
estimated $\frac{1}{N}\log_2 Z$ is about $1.802$.

Figs.~\ref{fig:FerG} and \ref{fig:FerGD} show simulation
results for $J = 1.25$, obtained from uniform sampling on the original factor 
graph and on the modified dual factor graph, respectively. 
From Fig.~\ref{fig:FerGD}, the 
estimated $\frac{1}{N}\log_2 Z$ is about $2.928$.

Note that, at low temperature, we observe much faster mixing with Monte Carlo 
methods on the modified dual factor 
graph. 
On the dual factor graph, convergence improves as $J$ increases 
(i.e., temperature decreases),
which is in sharp contrast to convergence on the original factor graph. 

\begin{figure}[t!]
\centering
\includegraphics[width=\linewidth]{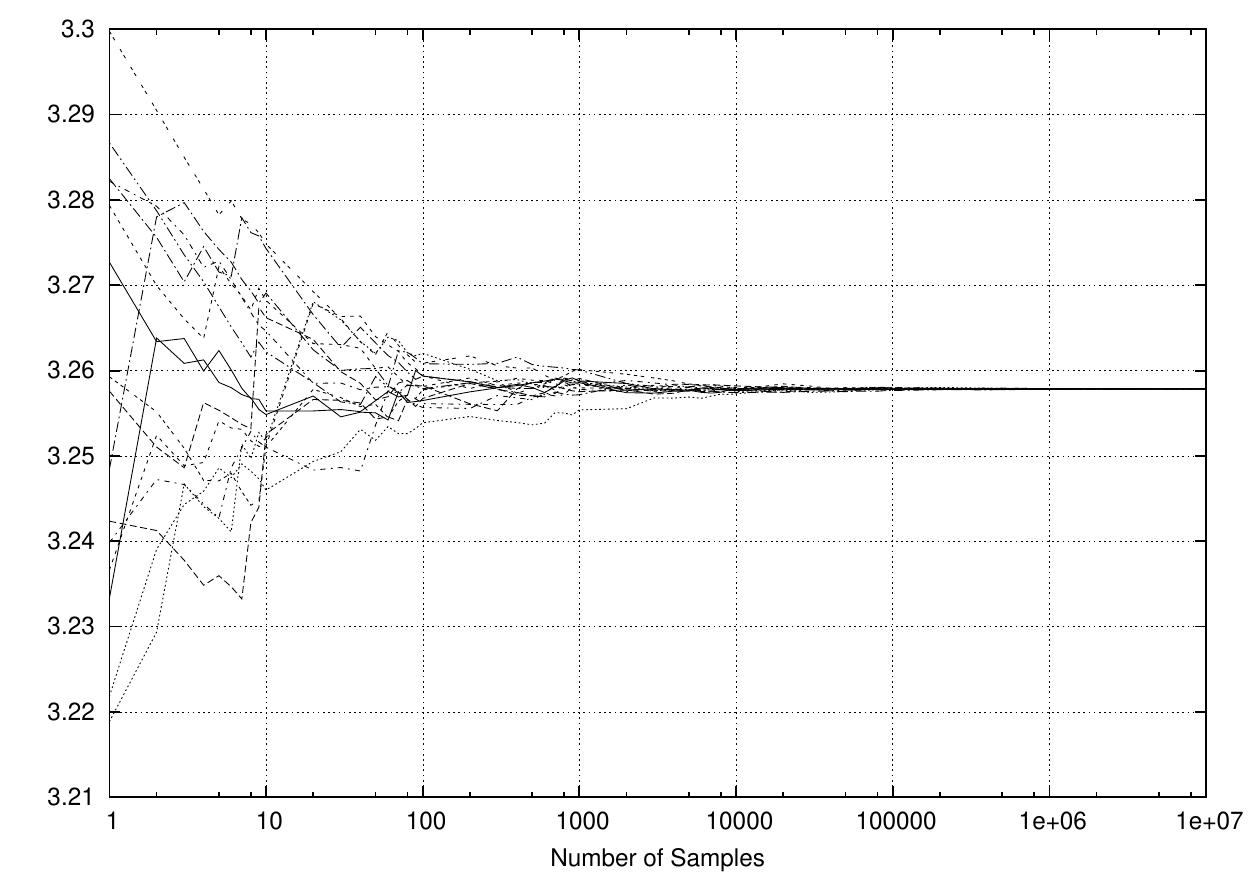}
\caption{\label{fig:SP1}%
Estimated free energy per site vs.\ the number of samples
for a $10\times 10$ ferromagnetic Ising model with 
$J\sim\calU[1.0, 1.5]$ using uniform sampling on the modified dual factor graph.
The plot shows 15 different sample paths.}
\end{figure}

\subsection{2D Ising model with spatially varying couplings}
\label{sec:2DSpin}

We apply uniform sampling on the modified dual factor graph
to estimate $\frac{1}{N}\log_2 Z$ of 
2D ferromagnetic Ising models with spatially varying couplings.
In our experiments, we set $J_{k, \ell} \sim \calU [1.0,1.5]$ 
independently for each factor. 

\begin{figure}[t!]
\centering
\includegraphics[width=\linewidth]{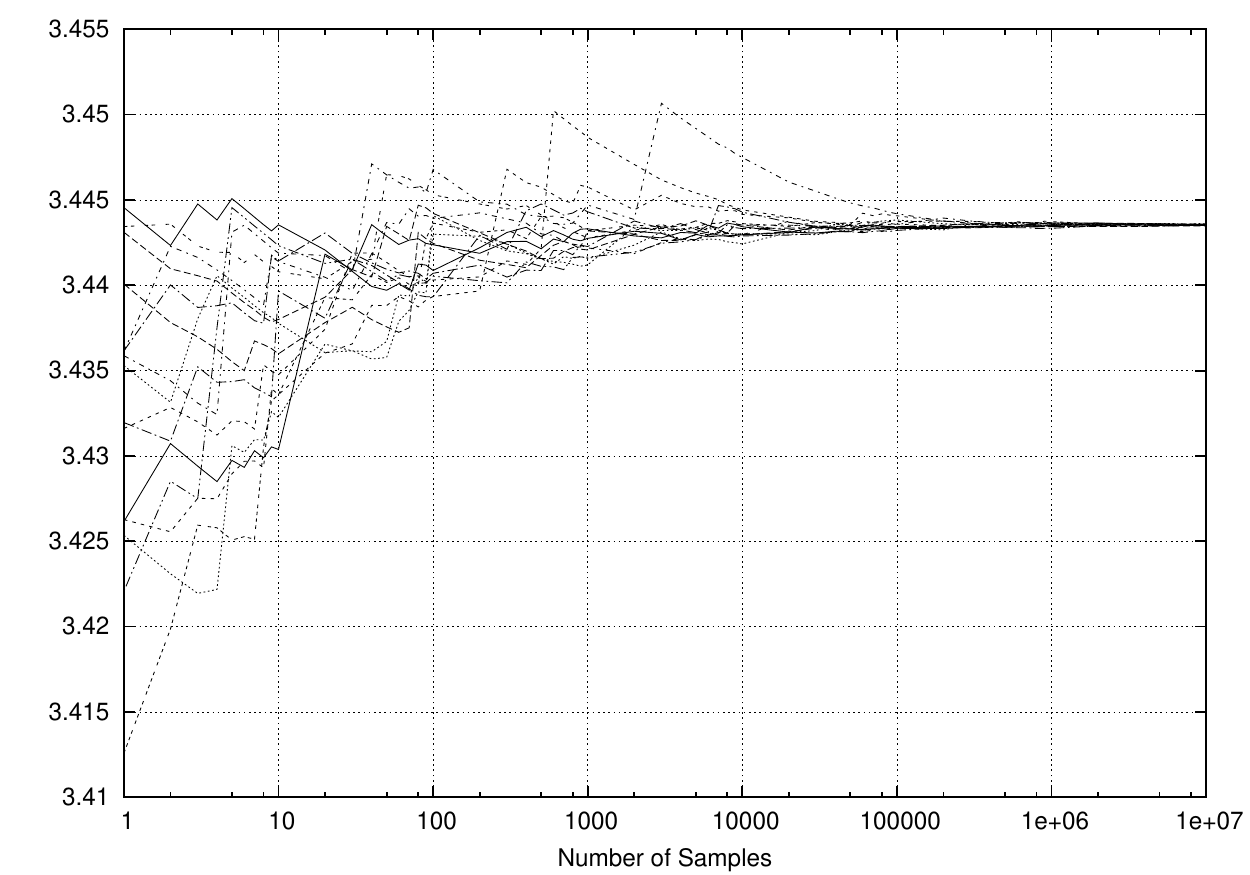}
\caption{\label{fig:SP2}%
Everything as in Fig.~\ref{fig:SP1}, but 
with $N = 20\times 20$.}
\end{figure}

For $N = 10\times 10$,
\Fig{fig:SP1} shows simulation results 
for one instance of the Ising model, where the 
estimated $\frac{1}{N}\log_2 Z$ is about $3.258$.
For one instance of such an Ising model with size 
$N = 20\times 20$, simulation results for 
$\frac{1}{N}\log_2 Z$ are shown  
in~\Fig{fig:SP2}. The 
estimated free energy per site is about $3.443$. 

As in our numerical experiments in Section~\ref{sec:2DConstant}, we observe 
fast mixing using uniform sampling on the modified dual factor graph.
Convergence of Monte Carlo methods improves as $J$ increases 
(or equivalently as the temperature $T$ decreases). 

\section{Conclusion}

The dual factor graph theorem~\cite{Forney:01, AY:2011, FV:2011} offers new possibilities for 
computing the partition function of Ising models. In particular, Monte
Carlo methods work much better on the dual graph, especially at low 
temperature.
A comparison with the Swensen-Wang algorithm~\cite{SW:87} needs to 
be addressed in future work.
Also, the relation between factor graph duality and 
the Kramers-Wannier duality~\cite{KW:41} should be
investigated.



\newcommand{\IT}{IEEE Trans.\ Inf.\ Theory}
\newcommand{\CASI}{IEEE Trans.\ Circuits \& Systems~I}
\newcommand{\COM}{IEEE Trans.\ Comm.}
\newcommand{\COMLet}{IEEE Commun.\ Lett.}
\newcommand{\COMMag}{IEEE Communications Mag.}
\newcommand{\ETT}{Europ.\ Trans.\ Telecomm.}
\newcommand{\SPMag}{IEEE Signal Proc.\ Mag.}
\newcommand{\ProcIEEE}{Proceedings of the IEEE}

\end{document}